# Directional Dependence of the Electronic and Transport Properties of 2D Monolayer Orthorhombic Diboron Dinitride (o-B$_2$N$_2$): DFT coupled with NEGF Study

Rameshwar L. Kumawat[†], Biswarup Pathak*[,†,#]

[†]Department of Metallurgy Engineering and Materials Science, and [#]Department of Chemistry, Indian Institute of Technology (IIT) Indore, Indore, Madhya Pradesh, 453552, India
*E-mail: biswarup@iiti.ac.in

## ABSTRACT

Tuning two-dimensional (2D) nanomaterial's structural and electronic properties has facilitated the new research paradigm in electronic device applications. In this work, the first-principles density functional theory (DFT) based methods are used to investigate the structural, electronic, and transport properties of an orthorhombic diboron dinitride (o-B$_2$N$_2$)-based polymorph. Interestingly, it depicts a low band gap semiconducting nature with a robust anisotropic behaviour compared to the hexagonal boron nitride (h-BN), which is an insulator and isotropic. We can also tune the structural and electronic properties of the semiconducting BN-based structure through an external in-plane mechanical strain. Further, by employing the Landauer-Büttiker approach, the electronic transmission function [$T(E)$], and electric current ($I$) calculations reveal that the boron nitride-based polymorph shows a robust direction-dependent anisotropy of the quantum transport properties. We have demonstrated the direction-dependence of the electric current in two perpendicular directions ($I_x$ and $I_y$), where we have observed an electric current ratio ($\eta = I_x/I_y$) of around 61.75 at 0.8 V. All these findings, such as directional-dependence anisotropy in [$T(E)$], I−V characteristics, and bandgap tunning, suggest that the applicability of such B$_2$N$_2$ based monolayer can be promising for futuristic electronic device applications.

## 1. INTRODUCTION

In condensed-matter physics and materials science, semiconductors are the basis of several important nanotechnologies like electronics, optoelectronics, sensing, computing, and communications.[1–3] In 1947, John Bardeen and Walter Brattain had succeeded transistor action in a germanium point contact nanodevice. This experiment paved the way for developing several distinct and integrated semiconductor nanodevices and nanocircuits.[4] Hence, in recent years, low-dimensional semiconductor nanomaterials are becoming omnipresent components of quotidian life[1]. The electronic bandgap is the critical factor that triggers the electrical and optical properties of these semiconductors.[1–3,5–7] The success in graphene development has motivated researchers and scientists to explore a new variety of two-dimensional (2D) nanomaterials beyond graphene, which holds semiconducting electronic bandgap.[2,3,8–11] The 2D nanomaterials show tunable electronic bandgaps properties, which can be achieved through van der Waals heterostructures,[12] atomic doping,[13] layers controlling,[14–16] intercalation,[17] strain engineering,[6] and so on.[5–7] 2D nanomaterials like bilayer graphene and black phosphorus (BP) show the electronic bandgap between terahertz and mid-infrared regions.[1,18] Transition metal dichalcogenides (TMDCs) exhibits the bandgap in the visible region,[1,19] whereas hexagonal boron nitride (h-BN) displays a bandgap in the ultra-violet region of the spectrum.[1,20–22]

2D h-BN is another class of layered materials analogous to graphite with a stable monolayer form identified as white graphene.[20] It is an insulator with a high bandgap (~6 eV) that attracted considerable attention for electronic device applications.[20–22] Nevertheless, due to the high electronic bandgap of h-BN, it could not be utilized properly. Tuning the electronic properties of such materials requires high mechanical strain, which results in structure deformation.[23] Therefore, the researcher has theoretically proposed another BN allotrope with N-N and B-B bonds. Demirci and co-workers have theoretically predicted a polymorph of BN monolayer in an orthorhombic structure (o-$B_2N_2$).[24] They have demonstrated that the structure is thermally and dynamically stable with a direct bandgap of 0.64 eV, which poses a different electronic structure compared to that of h-BN. Interestingly, the conduction and valence band edges of o-$B_2N_2$ are dominated by B ($p_z$) and N ($p_z$) orbitals, which can generate anisotropic nature in o-$B_2N_2$. This is unique in the sense that anisotropy is absent in h-BN. Therefore, it is very important to investigate the directional dependent structural, electronic, and anisotropic behavior of the o-$B_2N_2$ -based device. Using the density functional theory (DFT) calculations, we have investigated the structural, electronic, and anisotropic properties of the o-$B_2N_2$. Moreover, from the point of view of a device, we computed the electronic transmission function [$T(E)$], and electric current ($I$), which shows that o-$B_2N_2$ presents a strong directional anisotropic character of the transport properties. This o-$B_2N_2$ -based device offers an excellent and unique low-dimensional physics compared to isotropic 2D materials. Hence, it may open doors for directional dependent novel nanoscale device applications such as applications in optoelectronics, electronics, magneto-transport, anisotropic electrical transport, thermoelectric, ferroelectric, piezoelectric, and molecular detection.

Further, strain engineering objectives at altering the electronic bandgap properties of the 2D nanomaterials by applying an external strain. For example, strain engineering in silicon monolayer has allowed transistors to have four-times higher hole mobilities than unstrained nanodevices.[25] Furthermore, strain engineering in laser nanodevices enables vastly reduced threshold currents to be attained and allows accurate control over the emission wavelength.[26] Moreover, electronic, quantum transport, and optical properties are mostly determined by the atomic structure. It follows that the strain is the most suitable approach to tune the intrinsic properties of 2D nanomaterials, including graphene,[27] BP,[8,28] h-BN,[9,29] and TMDCs.[16] Experimentalists and theoreticians have shown that small strains induced by the substrate can be used to tune the structural and electronic properties of the systems. For example, strain in the graphene system reinforces the electron-phonon coupling.[8,27] In BP, the electronic and optical properties depend strongly on the applied strain.[28] First principal calculations have investigated the strain effect on the radiation hardness and sensing properties. Similarly, in TMDCs (for example, $MoS_2$), small compress/stretch strain increases/decreases the fluorescence properties and furthers it shows a semiconductor-metal-transition.[29–31] Therefore, we show that we can tune the structural and electronic properties of 2D semiconducting o-$B_2N_2$ nanomaterial through an external in-plane mechanical strain.

## 2. COMPUTATIONAL DETAILS

The geometrical and electronic structural calculations are done using the DFT methods, as implemented in the SIESTA code.[32,33] The GGA-PBE approximation is employed for exchange-correlation functional.[34] Troullier-Martins Norm-conserved pseudopotentials are applied to define the interaction between the core and valence electrons.[35] We have used a double-ζ-polarized (DZP) basis set to define the Kohn-Sham orbitals and a mesh cutoff of 500.0 Ry to describe the grid's density.[34] The Brillouin zone sampling is done with 16×32×1 $k$-points within the Monkhorst-Pack scheme. To minimize the interaction between the repeated images of the layers, a vacuum of 20 Å is used. All the structures are fully relaxed to obtain the minimum energy structure. The convergence-tolerance in the density matrix is $10^{-4}$ eV, and the residual forces on the atoms are kept less than $10^{-2}$ eV/Å.

We have investigated the electronic transport characteristics within the Landauer-B*ü*ttiker formalism using the non-equilibrium Green's function (NEGF) combined with DFT formalism, as implemented in the TranSIESTA code.[36] For this, we have used $1 \times 32 \times 32$ k-points. The exchange-correlation functional, basis sets, mesh cutoff, convergence tolerance, and residual forces are the same as described above for geometrical relaxations. The $T(E)$, which describes the probability of an electron to transfer from one electrode to the other with the energy ($E$), can be calculated as follows:

$$T(E) = \Gamma_L(E,V)\mathcal{G}(E,V)\Gamma_R(E,V)\mathcal{G}^\dagger(E,V) \qquad (1)$$

where the coupling-matrices are represented as $\Gamma_{L/R} = i[\Sigma_{L/R} - \Sigma_{L/R}^\dagger]$ and the NEGF for the device region represented as $\mathcal{G}(E,V) = [E \times S_s - H_s[\rho] - \Sigma_L(E,V) - \Sigma_R(E,V)]^{-1}$, where $S_s$ is the overlap-matrix and $H_s$ is the Hamiltonian-matrix, $\Sigma_{L/R} = V_{S_{L/R}} g_{L/R} V_{L/R} s$ is the self-interaction energy, $\Sigma_{L/R}$ is a lead (electrode) that takes into account from the L/R leads onto the device region, $g_{L/R}$ is the surface Green's function (L/R) and $V_{L/R} s = V_{S_{L/R}}^\dagger$ is the coupling-matrix between electrodes (L/R) and the device region.[36–43]

Moreover, the integration of $T(E)$ gives the $I(V_b)$ (where $V_b$ is applied bias voltage) as follows:

$$I(V_b) = \frac{2e}{h} \int_{\mu_R}^{\mu_L} T(E,V_b) [f(E - \mu_L) - f(E - \mu_R)] dE \qquad (2)$$

where $T(E, V_b)$ presents the transmission function of the electrons entering at $E$ from $L$ to $R$ electrode under $V_b$, $f(E - \mu_{L,R})$ is implying the Fermi-Dirac distribution function of e⁻ in the L/R electrodes and $\mu_{L,R}$ is the electrode chemical potential where $\mu_{L/R} = E_F \pm V_b/2$ are stimulated correspondingly up/down according to the Fermi-energy $E_F$.[2,3,36–40]

The applied strain in the system is calculated as $\varepsilon_x = (a - a_0)/a_0$, and $\varepsilon_y = (b - b_0)/b_0$, where $a(b)$ and $a_0(b_0)$ are the lattice parameters along the $x(y)$-direction for the strained and relaxed structures.

The strain energy (ΔE) in the system is calculated using the following equation:

$$\Delta E = E_{o-B_2N_2-strain} - E_{o-B_2N_2} \qquad (3)$$

Where $E_{o-B_2N_2-strain}$ and $E_{o-B_2N_2}$ are total energy of a fully relaxed system with/without strain.

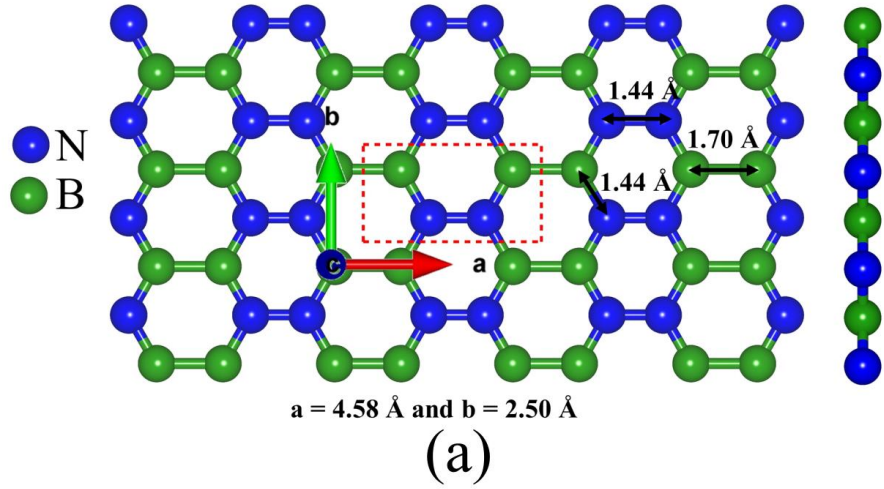

(a)

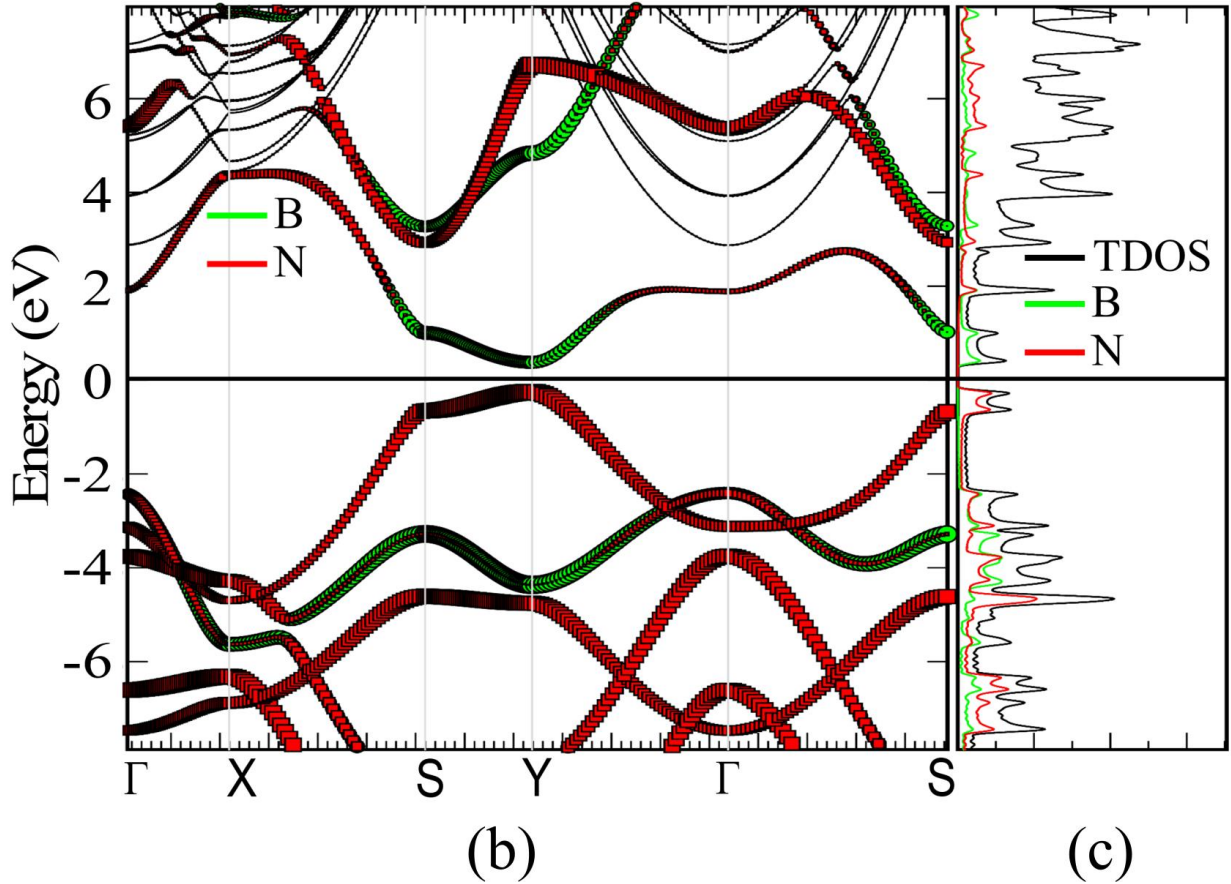

(b) (c)

**Figure 1**. (a) The ball-and-stick representation of the optimized o-$B_2N_2$ structure (top and side views). Here, the orthorhombic unit cell is red coloured. (b) The projected electronic band structures of o-$B_2N_2$, where red and green colours represent contributions of N and B atoms, respectively. (c) The electronic density of states (DOS) projected on B and N atoms. The Fermi-level is aligned to zero.

## 3. RESULTS AND DISCUSSION

*3.1 Structural and Electronic Properties:* The fully relaxed unit cell structure of 2D o-$B_2N_2$ is shown in **Figure 1(a)**. The calculated lattice parameters of the 2D orthorhombic unit cell are found to be a×b = 4.58×2.50 Å$^2$ with B-B, B-N, and N-N bond lengths of 1.70, 1.44, and 1.44 Å, respectively. Interestingly, the orthorhombic structure of the o-$B_2N_2$ unit cell has equal N-N and B-N bond length. Therefore, the B-B bond length is significantly longer compared to the N-N and B-N bond lengths. Thus, the o-$B_2N_2$ structure consists of a regular hexagonal structure. The projected electronic band structure and corresponding density of states (DOS) are shown in **Figure 1(b-c)**, shows a direct bandgap semiconductor with a bandgap of 0.64 eV at the Y-symmetric point, which agrees well with the previous report.[24] Further, we have made atomic projections of the band structure as shown in **Figure 1(b)**; one can see that both atoms (B and N) contribute to the valence band maximum (VBM) and conduction band minimum (CBM). Our analysis of the atomic orbital projections to the VB and CB edges are dominated by B-$p_z$ and N-$p_z$ atomic orbitals (**Figure S1; Supporting Information**), respectively. Further, we have checked the anisotropic character of the o-$B_2N_2$ monolayer. All these findings usher us to investigate further the directional-dependent structural, electronic, and transport properties of the o-$B_2N_2$ monolayer structure.

*3.2 Effect of Mechanical Strain on Structural and Electronic Properties:* We established that the structural and electronic properties could be tuned by using in-plane mechanical strain. Therefore, we have investigated the effect of mechanical strain on the structural and electronic properties of the o-$B_2N_2$ monolayer system. The uniaxial and biaxial mechanical strain along $x$- and $y$-directions is applied, which are mutually perpendicular to each other. First, we would like to discuss the effect of uniaxial mechanical strain in the $x$- and $y$-directions on the electronic strain energies (ΔE, in meV), bond lengths (Å), bandgap (Gap, in eV), and band structures of the o-$B_2N_2$ monolayer system.

*3.2.1 Uniaxial Mechanical Strain:* We first study the effect of uniaxial mechanical strain on the ΔE of o-$B_2N_2$ monolayer system. Starting from the fully optimized unit cell, the mechanical strain ranging from 1% to 6% is applied along $x$- and $y$-directions. A schematic of the uniaxial strain (compress/stretch in the $x$- and $y$-direction) is shown in **Figure 2**.

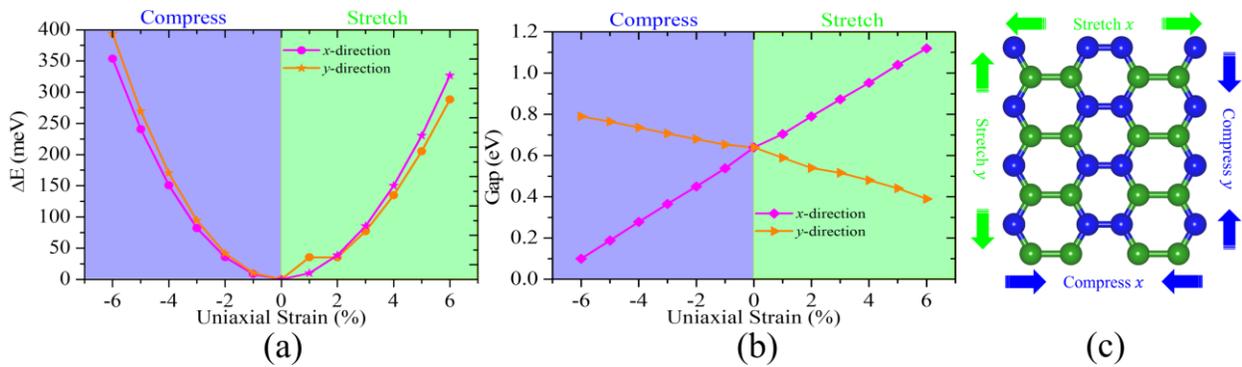

**Figure 2**. Effect of uniaxial compress/stretch strain on (a) electronic strain energy (ΔE, in meV). (b) Electronic bandgap (Gap, in eV) in the o-$B_2N_2$ system. (c) The schematic of o-$B_2N_2$ showing the compress/stretch direction.

As shown in **Figure 2(a),** it is noted that a compressive strain of -1% to -6% leads to change in the ΔE amounting from 8.41(10.29) to 353.71(393.69) meV in $x(y)$-direction. On the other hand, a stretch strain of 1% to 6% also leads to change the ΔE from 35.42(9.92) to 288.42(326.73) meV in $x(y)$-direction as shown in **Figure 2(a)**. Thus, the uniaxial strain significantly affects the ΔE values of the o-$B_2N_2$ monolayer system, and a parabolic curve has been observed as the applied strain (compressive/stretch) increases. We further studied the change in bond lengths (Å) under different types of mechanical strains. Under the uniaxial compressive strain along x(y)-direction, the bond length decreases monotonously with increasing strain [**Figure S2(a)**], while with increasing uniaxial stretch strain along x(y)-direction, the bond length shows a linear increase (**Figure S2(b)**). The maximum bond length reduces of 0.1, 0.03, and 0.06 Å for B-B, B-N, and N-N, respectively under uniaxial strain along the x-direction. While the bond length compression of 0.02, 0.05, and 0.04 Å is observed for B-B, B-N, and N-N bond, respectively, for the uniaxial strain along the y-direction. Similarly, the maximum bond length stretches of 0.15, 0.02, and 0.05 Å for B-B, B-N, and N-N is observed along the x-direction, whereas the bond length stretches of 0.03, 0.05, and 0.02 Å for B-B, B-N, and N-N is observed along the y-direction. Thus, we find that an increase in compressive/stretch strain leads to a decrease/increase in the bond length parameters. This is due to the application of an external strain (compressive/stretch), atomic positions of the lattice get distorted; hence we found these changes accordingly.

Further, we investigated the effect of compressive/stretch strain on the electronic band structure and bandgap properties. We find that the in-plane compressive/stretch strain significantly tunes the bandgap of the o-$B_2N_2$ monolayer system, as shown in **Figure 2(b)**. The compressive strain of 1% to 6% along the *x*-direction can reduce the bandgap from 0.64 eV to 0.1 eV, whereas the bandgap increases from 0.64 eV to 0.79 eV if the strain is applied along the y-direction (**Figure 2(b)**). Similarly, the stretch strain along the *x*-direction increases the band gap from 0.64 eV to 1.12 eV, whereas reduces the band gap (0.64 eV to 0.39 eV) while applied along the y-direction (**Figure 2(b)**). This happens because the lattice periodicity gets disturbs due to the application of an external in-plane uniaxial strain; hence, the VBM/CBM position shifts under strain. This leads to an increase/decrease in the electronic bandgap. Interestingly, the h-BN monolayer does not show many variations in the band gap under strain in the zigzag direction.[44] Therefore, we have found that the electronic properties of the o-$B_2N_2$ monolayer can be easily tuned compared to that with the h-BN monolayer.

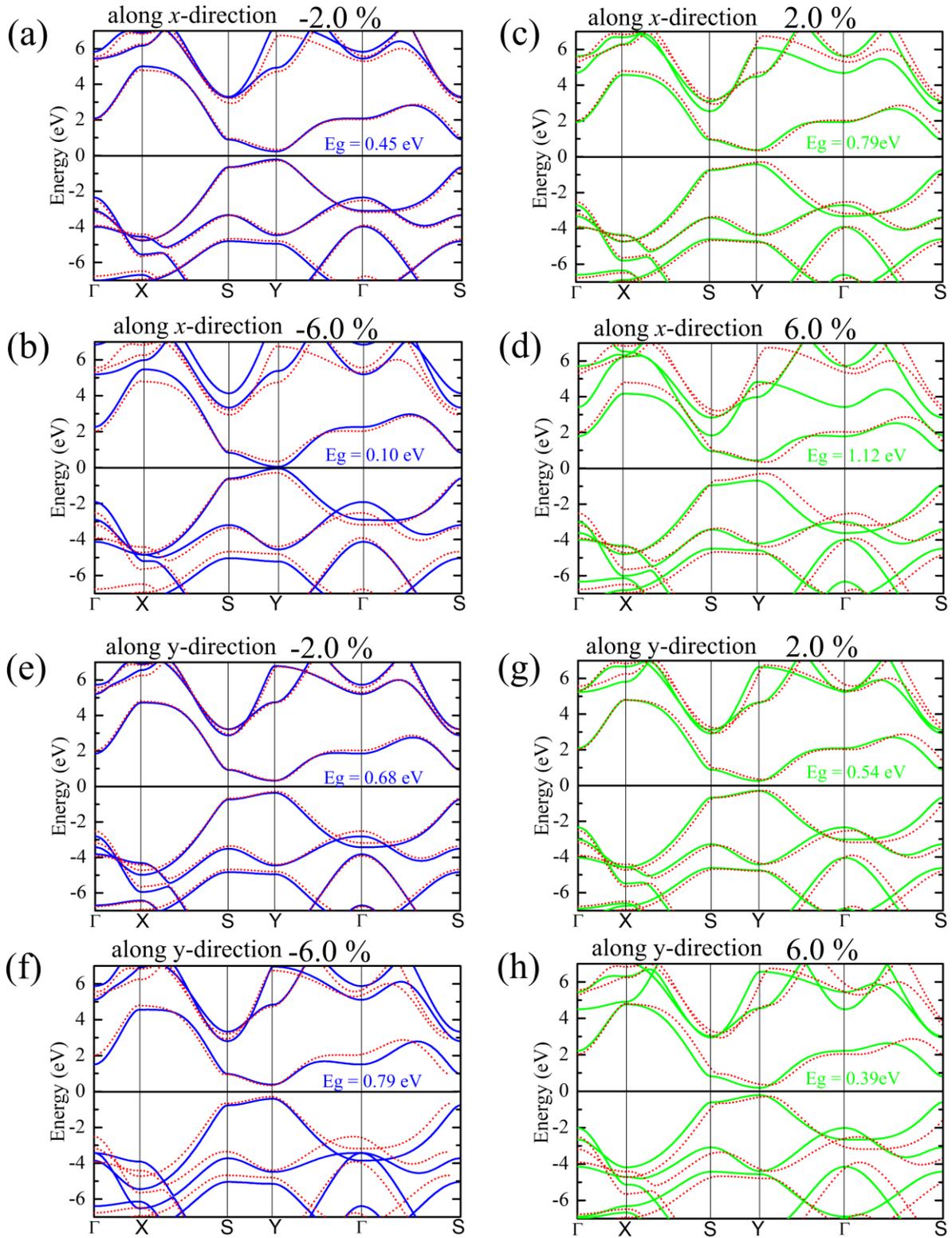

**Figure 3**. Electronic band structures of o-$B_2N_2$ under uniaxial compressive/tensile strains (2 and 6 %) along x- and y-directions.

Further, in **Figure 3,** we show the electronic band structures o-$B_2N_2$ monolayer system under compressive/tensile strains along the x- and y-directions. The compressive strains of 2% and 6% reduce the electronic bandgap (from 0.64 eV to 0.45 and 0.10 eV) at the Y-symmetric point, as shown in **Figure 3**. On the other hand, the stretching strains of 2% and 6% increase the electronic bandgap (from 0.64 eV to 0.79 eV at 2% and to 1.12 eV at 6%) at the Y-symmetric point. Similarly, we find that 2% and 6% compressive strain in the $y$-direction also affects the electronic band structure at the Y-symmetric point, but the band movement is not significant [**Figure 3(e-f)**]. Interestingly at the Y-symmetric point, the CBM remains unperturbed, whereas the VBM is pushed down. The 2% and 6% stretch strain in the $y$-direction reduces the bandgap (from 0.64 eV to 0.54 eV at 2% and 0.64 to 0.39 eV at 6%) at the Y-symmetric point, and the CBM and VBM are moving towards the Fermi energy. Thus, the strain along the $y$-direction has the opposite effect compared to that in the $x$-direction.

*3.2.2 Biaxial Mechanical Strain:* We next investigate the effect of biaxial mechanical strain on the electronic ΔE (in eV), bond length (Å), bandgap (Gap, in eV), and band structures of the o-$B_2N_2$ monolayers system. **Figure 4** shows the schematic of the biaxial strain (compress/stretch) in the xy-direction. From **Figure 4(a)**, we note that a compressive strain of -1% to -6% leads to change in the ΔE values amounting from 21.84 to 917.43 meV, while a stretch strain of 1% to 6% leads to change in the ΔE values from 21.81 to 672.97 meV. Therefore, the biaxial strain significantly affects the ΔE values of the o-$B_2N_2$ monolayer system. A parabolic curve can be observed as we plot energy change vs. applied strain (compress/stretch).

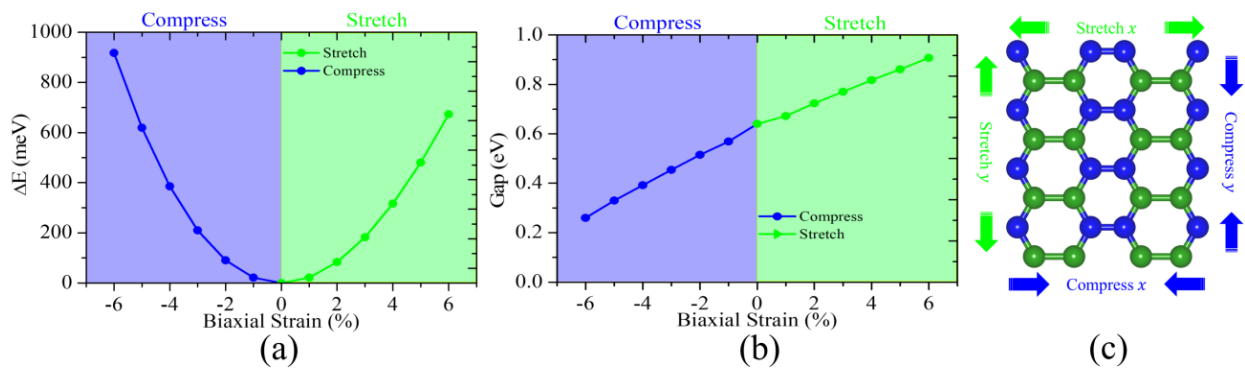

**Figure 4**. Effect of bidirectional compress/stretch strain on (a) electronic strain energy (ΔE, in meV). (b) Electronic bandgap (Gap, in eV) in the o-$B_2N_2$ system. (c) The schematic of o-$B_2N_2$ showing the compress/stretch direction.

We further studied the change in bond lengths (Å) under biaxial strain. Under the biaxial compress strain along the $x$y-direction, the bond length decreases monotonously with increasing strain [**Figure S3**]. Similarly, with increasing biaxial stretch strain along the $x$y-direction, the bond length increases. The maximum bond length reduces by 0.13, 0.08, and 0.09 Å for B-B, B-N, and N-N, respectively. While the bond length stretching of 0.16, 0.07, and 0.09 Å is observed for B-B, B-N, and N-N, respectively. Thus, we find an increase in biaxial compressive/stretch strain leads to a decrease/increase in the bond length parameters.

**Figure 4(b)** shows the bandgap as a function of biaxially applied mechanical strain. When the biaxial compressive strains are applied, the electronic bandgap decreases with increasing the mechanical strain up to 6%. The electronic bandgap decreases to a minimum of 0.26 eV at a compressive strain region of 6%. In the biaxial stretch strains region, the electronic bandgap increases with increasing the strain up to 6%. We note that the electronic bandgap rises to a maximum of 0.9 eV at a strain of 6%. In contrast, the electronic bandgap of h-BN decreases as the biaxial stretch strain increases, and it increases/decreases as the biaxial compressive strain increases up to 5%.

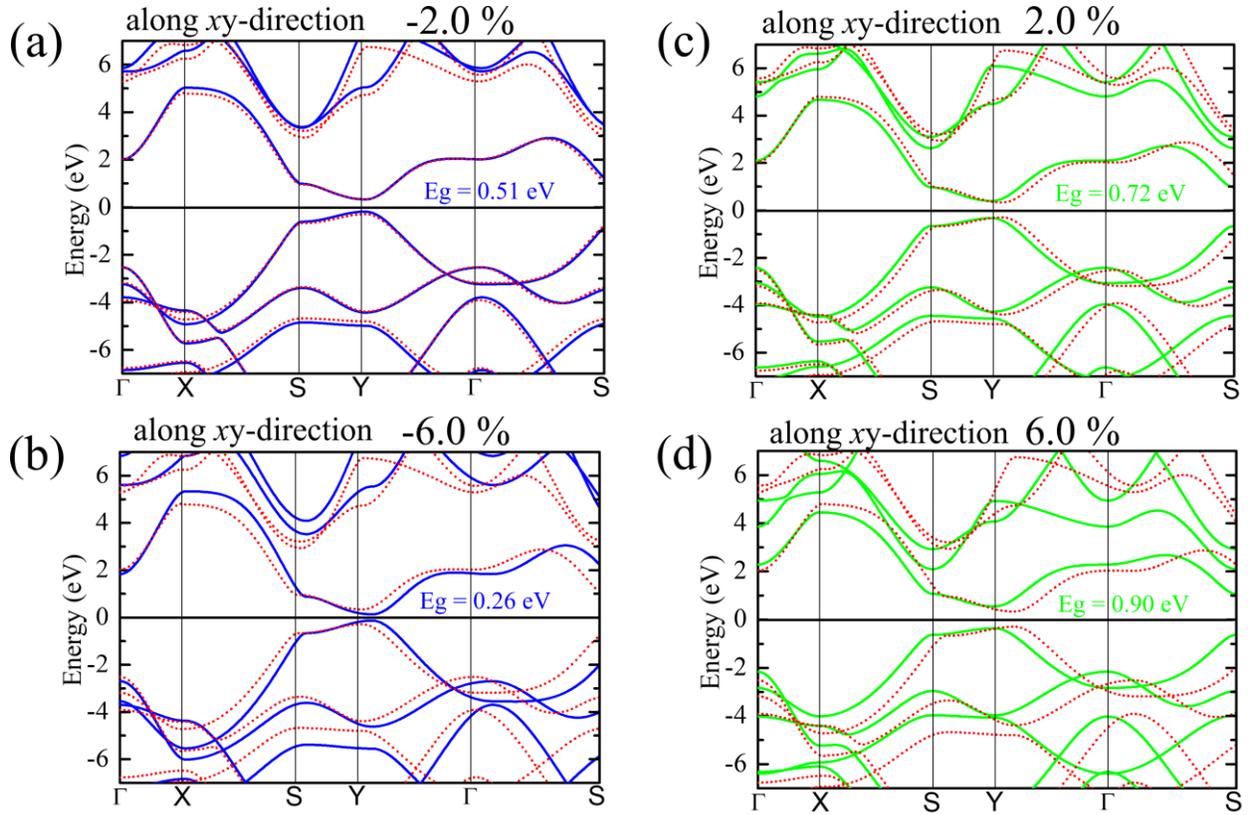

**Figure 5**. Electronic band structure for bidirectional compress [a-b; left side] and stretch (c–d; right side) in the o-$B_2N_2$ monolayer system.

We have also examined the electronic band structure of the compress/stretch strained o-$B_2N_2$ monolayers system. **Figure 5(a-b)** and **5(c-d)** show the electronic band structure of 2%, and 6% compressive/stretch strained o-$B_2N_2$ monolayers, respectively. When the 2% compressive strain is applied, the CBM remains constrained located at the Y-symmetric point, while the VBM, which is located at the Y-symmetric point, moves downward compared with that of the unstrained o-$B_2N_2$ monolayer (**Figure 5(a)**). Similarly, when the 6% compressive strain is applied, the CBM moves downward, and the VBM moves upward, located at the Y-symmetric point, as shown in **Figure 5(b)**. In short, the CBM and VBM moving towards the Fermi energy at the Y-symmetric point. On the other hand, the 2% and 6% biaxial stretch strain increases monolayer's bandgap, and the CBM moves upward while the VBM remains constrained located at the Y-symmetric point, as shown in **Figure 5(c-d)**. Thus, we find that under applied mechanical strain, the o-$B_2N_2$ monolayer remains a direct bandgap semiconductor. This makes o-$B_2N_2$ a very suitable material for electronic and optoelectronic applications such as other direct bandgap 2D materials, including $WS_2$, $MoSe_2$, and $MoS_2$.

*3.3 Electronic Transport Properties:* Understanding the intrinsic electronic transport properties helps us to understand the solid-state nanomaterials capability for nanoscale electronic device applications. Therefore, we have investigated the transport properties of the semiconducting o-$B_2N_2$ monolayer. A schematic illustration of the proposed nanoscale setup used for the electronic transport calculations along the $x(y)$-direction is shown in **Figure 6**. This setup consists of two semiconducting electrodes [left (L)/right (R)] composed of the semi-infinite o-$B_2N_2$ sheet. Both L/R electrodes are connected to the central scattering region, as shown in Figure 2. To confirm the directional-dependence on the transport properties, we have evaluated the transmission $T(E)$ along the $x$ and $y$-directions [$T_x(E)$ and $T_y(E)$]. The calculated $T(E)$ along the $x$ and $y$-direction are shown in **Figure 6(b)**. A step-like $T(E)$ curve has been observed. This means that integer $T(E)$ values are corresponding to the number of conduction channels. Moreover, it is observed that integer $T(E)$ values and patterns are different in $x(y)$ direction. In $x$-direction, we have found that the maximum value of $T_x(E)$ is reached up to 0.22 channels/Å. It has a higher transmission below the Fermi-level (at E-$E_F$ = -2.5 eV is 0.22 channels/Å) and a lower transmission value above the Fermi-level (at E-$E_F$ = 2.5 eV is 0.16 channels/Å). On the other hand, in $y$-direction, it has been found that $T_y(E)$ has a higher transmission value above the Fermi-level (0.31 channels/Å at E-$E_F$ = 2.5 eV) and lower transmission value below the Fermi-level (0.12 channels/Å at E-$E_F$ = -2.5 eV). All these can be explained from the band structure around the Fermi level, as there is a gap along the Γ-X path, whereas along the Y- Γ path, there are bands above the Fermi-level, which lead to higher transmission values for $T_y(E)$. These results reveal that the transmitting wave practically does face the scattering regions within a given $E$ window in both directions.

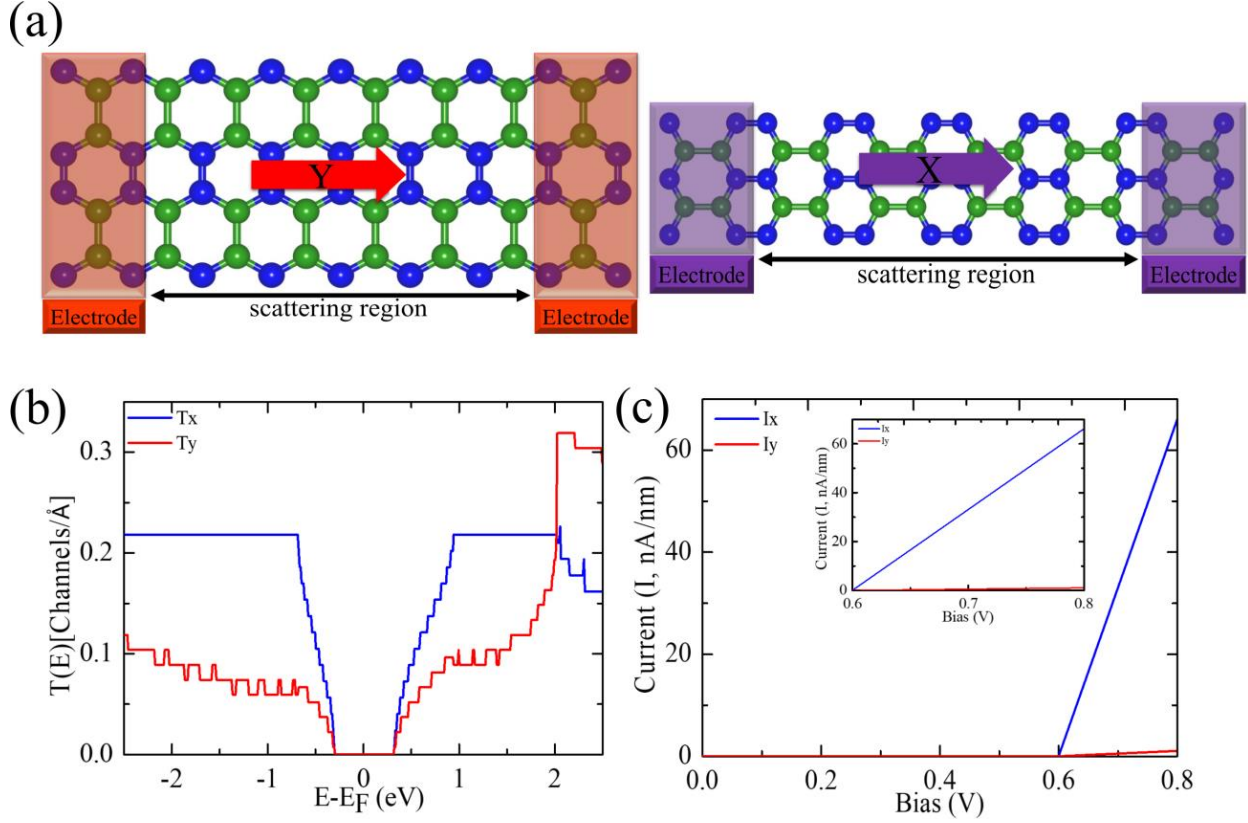

**Figure 6.** (a) Schematic illustration of the proposed setup utilized for transport calculations indicating the $x(y)$-directions. (b) The $T(E)$ for the $x$ ($T_x$) and $y$ ($T_y$)-directions. (c) The electronic current under $V_b$ in the $x$ ($I_x$) and $y$ ($I_y$)-directions.

Further, we have increased the bias voltage (V) in a stepwise manner using the DFT+NEGF approach. The Landauer-Büttiker method is used to calculate the electric current ($I$) along the $x$- and $y$-directions [as $x(I_x)$ and $y(I_y)$] of the o-$B_2N_2$ device. The electric current is calculated by performing the $T(E)$ integration in either direction by using equation 2. **Figure 6(c)** represents the I-V characteristics against different bias voltages. The I-V characteristics curve shows zero current up to 0.6 V, which is because of the semiconducting nature of the electrode with an electronic bandgap of 0.64 eV. Therefore, a minimum bias voltage of 0.64 V is required for electric current to flow through the device region. As the applied bias voltage increases, from 0.64 to 0.8, a rapid increase in the electric current can be seen along the x-direction. The *I-V* characteristics curve [the inset of **Figure 6(c)**] shows anisotropic nature in the electric current, and the calculated anisotropy ($\eta$) is found to be 61.75 ($\eta = I_x/I_y$) at 0.8 V. We have further compared our findings with other reported 2D-based structures. We note that the o-$B_2N_2$-based device shows relatively high anisotropy when compared to borophane, borophene, and BP-based devices.[2,3,45] This indicates that o-$B_2N_2$ -based nanoscale device holds excellent possibilities for different applications. This anisotropy may offer different properties along different directions caused by the atomic arrangement of the o-$B_2N_2$ system. On this basis, the BN-based system may be promising for directional-dependent optoelectronics, electronics, magneto-transport, anisotropic electrical transport, thermoelectric, piezoelectric, and molecular detection-based applications.

There are also possibilities to tune the anisotropy factor by increasing the device length, which will imply a quantum well, as discussed by Zhang and co-workers. Further, one can take advantage of the anisotropic behavior of the o-$B_2N_2$ nanostructure by making bilayer nanojunction and twisting them. This may open the possibility of o-$B_2N_2$ to use as a field-effect transistor in electronic devices.

## 4. CONCLUSIONS:

In conclusion, we have demonstrated that the 2D o-$B_2N_2$ monolayer structure exhibits robust structural, electronic, and transport properties. Using DFT+NEGF calculations, we show that it has a semiconducting band structure with a robust anisotropic behavior. We have shown that the bandgap properties can be tuned by applying uniaxial/biaxial in-plane mechanical strain. It is found that a compressive/stretch strain of 1% to 6% leads to a change in the electronic strain energies (ΔE, in meV), bond lengths (Å), bandgap (Gap, in eV), and band structures of the o-$B_2N_2$ monolayer system. It has been noted that a uniaxial/biaxial strain significantly affects the ΔE values of the o-$B_2N_2$ monolayer system, and a parabolic curve has been observed as the applied strain (compressive/stretch) increases. We find that an increase in uniaxial compressive/stretch strain leads to a decrease/increase in the bond length parameters. Also, the in-plane compress/stretch strain significantly modulated the electronic bandgap of the o-$B_2N_2$ monolayer system. An increase in biaxial compressive/stretch strain leads to a decrease/increase in the bond length parameters. The electronic bandgap rises a maximum of 0.9 eV at a biaxial strain region of 6%. In contrast, it has reported that the electronic bandgap of h-BN decreases as the biaxial stretch strain increases, and it increases/decreases as the biaxial compressive strain increases up to 5%. Further, the transmission function and electric current results reveal a strong direction-dependent anisotropy of the quantum transport properties. We find an electric current ratio ($\eta = I_x/I_y$) of around 61.75 at 0.8 V. Thus, all these findings suggest the applicability of the 2D o-$B_2N_2$ structure for futuristic electronic device applications.

## 5. ASSOCIATED CONTENT

**Supporting Information*

Projected density of states plot, the effect of uniaxial/biaxial mechanical strain on bond length parameters.

## 6. ACKNOWLEDGMENTS

We acknowledge IIT Indore for the lab and computing facilities. This work is supported by SPARC/2018-2019/P116/SL and DST-SERB (Project Number: CRG/2018/001131). R.L.K.

acknowledges MHRD for senior research fellowship (SRF). We would like to thank Dr. Vivekanand Shukla for fruitful discussion throughout this work.

## 7. DATA AVAILABILITY

The data that support the findings of this study are available from the corresponding author upon reasonable request.

## 8. CONFLICTS OF INTEREST

No conflicts of interest to declare.

## 9. ORCID

Rameshwar L. Kumawat: 0000-0002-2210-3428

Biswarup Pathak: 0000-0002-9972-9947